\documentclass[a4paper,11pt]{article}
\usepackage{fullpage}
\usepackage{amsmath}
\usepackage{amssymb}
\usepackage[dvips]{graphicx}
\usepackage{natbib}
\usepackage{setspace}
\usepackage{lineno}

\usepackage{xcolor,calc}

\title{\textbf{Analysis of a capture-recapture estimator for the size of populations with heterogenous catchability, and its evaluation on RDS data from rural Uganda}}
\author{Yakir Berchenko\\University of Cambridge,\\Department of Veterinary Medicine \\Madingley Road, Cambridge CB3 0ES,\\United Kingdom\\{\tt yb236@cam.ac.uk}
\and Richard G. White\\ Centre for the Mathematical Modelling of Infectious Disease\\ and Department of Infectious Disease Epidemiology\\
Faculty of Epidemiology \& Population Health\\ London School of Hygiene and Tropical Medicine\\
Keppel Street,
London,
WC1E 7HT\\
{\tt richard.white@lshtm.ac.uk}
\and Cyprian Wejnert\\Cornell University, Ithaca, NY
\and Simon D.W. Frost\\University of Cambridge,\\Department of Veterinary Medicine \\Madingley Road, Cambridge CB3 0ES,\\United Kingdom\\{\tt sdf22@cam.ac.uk}}

\begin{document}

\maketitle


\begin{abstract}
In this paper, we consider capture-recapture experiments with heterogenous catchability. In the setting we consider, the widespread Huggins-Alho estimator is not very suitable and we introduce and study a new generalized Horvitz-Thompson estimator. Our motivation is Respondent Driven Sampling (RDS), a prime example for such a setting where the capture probability is dependent on both the unknown population size as well as on an observable covariate, the network degree of an individual, due to peer recruitment. After discussing the theoretical properties of the new estimator, with full details given in the appendix, we evaluate it on various empirical and simulated data-sets, focusing on an RDS
survey in a population in rural Uganda in which the population size is  known \textit{a priori}. The results thus obtained demonstrate that the adjusted estimator is less biased than the naive Lincoln-Petersen estimator.

\end{abstract}

\section{Introduction}

When using naive capture-recapture methods to estimate the size of a population under investigation, it is well known that if the population sampling is not homogeneous (i.e., heterogeneous catchability) the estimates can be biased.
One way to overcome this problem is to employ models in which detection probabilities are viewed
as latent parameters described by some probability distribution \cite{NorrisPollock1996, Royle06} (for a review and criticism of this approach, see \cite{Link03}).

An alternative approach, which is more relevant to the circumstances we are interested in, is to
model heterogeneity explicitly by  identification of individual covariates that are thought to explain variation in detection probability (e.g., size, weight and age, which are commonly used in wildlife biology).
For instance, an early paper by Pollock, Hines and Nichols \cite{PHN} put forward a strategy of stratification of individuals into a finite number of classes, yielding $K$ strata with stratum population sizes $\{N_i\}^K_{i=1}$, where the collection of $N_i$ parameters is the object of inference. One obvious shortcoming of this method (in addition to the increase in parameter dimensions with the cardinality of the covariate) is that the covariates for the uncaptured animals are not observable and therefore,  Pollock et al had to categorize all animals into several groups and use the midpoint of the classifying covariate as a representative covariate for the group. Another strategy \cite{HUggins1989,Alho1990} avoids this difficulty through linear logistic modelling of capture probabilities, using the individual covariates of the captured individuals, thus obviating the need for the covariates of the uncaptured in the analysis.
In this strategy, $N$ is a derived
parameter, its estimation is based on a generalized Horvitz-Thompson estimator (HT), and classical methods of asymptotic inference are employed (for a comprehensive review and comparison with a third approach, the so-called \textquotedblleft joint likelihood" approach, which specifies the joint distribution of the capture data and the covariate, see \cite{Pollock02}).

Despite the justified widespread use of the Huggins-Alho estimator (HA), it is known to be problematic if capture probabilities are low (see \cite{BZF1998}, and also the limitation in  \cite{Alho1990} (p. 626) on the catchability  from having a large number of small values). In addition, as Royle points out \cite{Royle09}, the Huggins-Alho estimator has the conceptually unappealing aspect to it that $N$ (the object of inference)
is a derived parameter, formulated explicitly as a function of nuisance parameters (the capture probabilities, which are usually not of direct interest).

The above two shortcomings are particularly crucial for the case we are interested in here, where the capture probabilities depend on the size of the population, and hence can also take many small values. This contrasts with common scenarios in wildlife biology where, for example, the size of an individual has a simple bearing on its capture probability irrespective of $N$, and thus the HA estimator can sum the inverses of the (estimated) capture probabilities to produce an estimate of $N$.
One example for such a situation, where the capture probabilities depends on $N$, is respondent driven sampling (RDS) \cite{Heckathorn1997,Heckathorn2002}.

RDS is an approach to sampling design for \textquotedblleft hidden" populations such as marginalized or highly stigmatized populations (e.g., injection drug users, men who have sex with men, sex workers).
RDS utilizes the networks of social relationships that connect members of the target population to facilitate sampling by chain referral methods. Although this is expected to result in biased sampling (such as over-sampling participants with many acquaintances), most RDS studies, interested, for example, in the \textit{prevalence} of a disease within a population, typically conduct a \textquotedblleft single stage" survey and use the information about how members of the target population are connected to weight recruits in a way
to attempt to account for biased sampling \cite{Heckathorn1997,Heckathorn2002}. Although RDS studies typically have a single stage design, it is possible to use RDS data as the second stage of a two-stage survey, such as in capture-recapture or the \textquotedblleft multiplier method", with the aim of estimating the \textit{size} of a population (here we do not elaborate on the differences between the two methods since our approach applies to both. Briefly however, in a strict capture-recapture one has two random samples obtained by actively sampling twice.  In the  \textquotedblleft multiplier method", only one of the samples is random, such as the RDS sample, and the other sample need not be random, e.g.  service data).
Almost a decade has past since the first application of RDS in a capture-recapture setting  \cite{Heckathorn2002a, Heckathorn2003} with little justification or analysis since \cite{BerchenkoFrost}. Nevertheless, there has been recently a flourish of interest in this approach \cite{Curitiba11, PB,Russia10,Vietnam10}, with additional RDS surveys completed recently, or underway, in more than 15 countries as part of a two-stage survey designed to estimate the size of various hidden populations \footnote{For more detailed information contact the corresponding author}.


Due to peer recruitment, the sampling probabilities in RDS, $\{\pi_i\}^N_{i=1}$, are assumed to be proportional to the degree, i.e. $\pi_i = \frac{d_i}{zN}$ (where $d_i$ is the degree of individual $i$ and $z$ is the mean degree) though still unknown, because the size of the population is unknown.
Notice that although we may start with a more explicit form of the catchability than HA, here we do not go further and attempt to estimate it exactly; only a proportionality relation is required.

In this paper, we present the first detailed analysis of a generalized HT estimator which is highly applicable to cases such as those described in the previous paragraph. Although the sample design and the  assumptions are a bit different than those used in the HA approach, they are not more restrictive and can be extended to many other cases.

After introducing some notations and our new adjusted estimator in section 2.1,  we start by analyzing the behavior of a naive Lincoln-Petersen estimator (LP), before comparing it to the adjusted estimator. In section 2.2  we show that the adjusted estimator has the desirable properties of converging to true population size in probability as $N \rightarrow \infty$, as well as being indifferent to unknown capture probabilities in one of the two stages. In section 3 we test the adjusted estimator on various empirical data-sets, focusing mostly on data from an
RDS survey in a population in rural Uganda in which the population size is  known \textit{a priori} (sec. 3.1), but also in section 3.2 on two additional data-sets.
This can also be regarded as an examination of the utility of RDS sampling design and weighting, which have been under some consideration lately \cite{Goel2010, Gile11}.
Our main findings in section 3 are that RDS sampling and the adjusted estimator allow us to obtain a relatively accurate estimate of the population size (in contrast to the naive approach).
%
In section 4 we discuss our findings and the ways in which such an analysis could be applied easily and without violating confidentiality (of non-RDS individuals) in similar circumstances.

\section{Theory and simulations}

\subsection{Setting and notations}
Let $N$ be the (unknown) size of the population to be estimated. After sampling the population twice independently, imagine that a researcher finds $a_{1,1}$ individuals who appeared in both samples,  $a_{1,0}$ individuals who appeared only in the first sample and  $a_{0,1}$ individuals who appeared only in the second sample. A naive estimate might be obtained by the Lincoln-Petersen estimator:
\begin{equation}
\hat{N}_{naive} = \frac{(a_{1,0}+a_{1,1})(a_{0,1}+a_{1,1})}{a_{1,1}}
\end{equation}
Unfortunately, even if the two samples are independent, unequal and positively correlated catchability in both sampling stages results in  an overrepresentation of \textquotedblleft marked" individuals in the sample and as a consequence a smaller $\hat{N}$ \cite{HRR, KMT}; this situation may easily happen, for example, if there are two (independent) RDS studies, or a \textquotedblleft sampling stage" comprised of records of people associated with a transmissible disease (where, in theory, the more friends you have the more likely you are of being infected).
A few papers \cite{Heckathorn2002a, Heckathorn2003, PB} have taken a different ad-hoc and heuristic approach. Taking an analogous assumption to the standard capture-recapture assumption, they assumed that the (weighted) proportion of individuals captured a second time (compared to the weighted size of the second sample) is equal to the proportion of individuals captured in the first sample (compared to $N$), that is:
\begin{equation}
\frac{\sum_{k}\frac{\mathbb{I}^{1,1}_{k}}{\pi_k}}{\sum_{k}\frac{\mathbb{I}^{1,1}_{k}}{\pi_k}+\sum_{k}\frac{\mathbb{I}^{0,1}_{k}}{\pi_k}} = \frac{\sum_{k} {\mathbb{I}^{1,1}_{k}}+\sum_{k}{\mathbb{I}^{1,0}_{k}}}{N}
\end{equation}
where $\mathbb{I}^{i,j}_{k}$ is the indicator whether individual $k$ was caught or not in each sample, and $\pi_k$ is his catchability. Rearranging (2) we get the following estimator for $N$:
 \begin{equation}
\hat{N}_{adj} = (\sum_{k}{\mathbb{I}^{1,0}_{k}}+\sum_{k}{\mathbb{I}^{1,1}_{k}})\frac{(\sum_{k} \frac{\mathbb{I}^{1,1}_{k}}{\pi_k}+\sum_{k}\frac{\mathbb{I}^{0,1}_{k}}{\pi_k})}{\sum_{k}\frac{\mathbb{I}^{1,1}_{k}}{\pi_k}}
\end{equation}
If the $\pi_i$'s are not known, but we are able to measure the value of a covariate, $\tilde{\pi}_i$, proportional to $\pi$, we can easily use it instead for (3). The case where the catchabilities in both samples are uncorrelated, in particular when the sampling in the second stage is uniformly at random, is well known to be trivial and similar to the case of equal catchability \cite{HRR, KMT}. Therefore we now focus on the case they are positively correlated, and in particular, using RDS as an example (where in both stages the probability that individual $k$ will be sampled is proportional to his degree $d_k$)
we start by showing why $\hat{N}_{naive}$ is a good estimator (in expectation) when the catchability is more or less homogeneous, and why it is not to be trusted when it is heterogeneous.

Assume in the first stage (second stage) individuals are sampled until $S_1$ individuals ($S_2$) are caught with

\begin{center}
$S_1 = a_{1,0}+a_{1,1} = \alpha_1 N $\\
$S_2 = a_{0,1}+a_{1,1} = \alpha_2 N $
\end{center}
Taking the RDS assumption (recruitment is performed in a \textquotedblleft random walk"-like manner, resulting with the degree-dependent catchability equal to the stationary distribution of a random walk on the network) with
$\frac{n \alpha_1}{z}$ as the correctly normalized probability for a person with degree $n$ to be caught in the first sample (similarly, $\frac{n \alpha_2}{z}$ is the probability of capture in the second sample), the expected number of individuals from a population with a degree distribution $\{p_n\}$ and mean degree $z:=\sum_n p_n n$ that are caught twice is:
\begin{equation}
\mathbb{E}[\sum_{k}\mathbb{I}^{1,1}_{k}] = \sum^{N}_{k=1}\sum_n p_n  \frac{n \alpha_1}{z} \frac{n \alpha_2}{z}
\end{equation}
Thus,
\begin{equation}
\mathbb{E}[\sum_{k}\mathbb{I}^{1,1}_{k}] = \frac{N\alpha_1 \alpha_2 m_2}{z^2}
\end{equation}
where $m_2$ is the second moment of the degree distribution. Before evaluating $\mathbb{E}[\frac {S_1 S_2} {\sum_{k}\mathbb{I}^{1,1}_{k}}]$, it is worth pointing out that in general $\sum_{k}\mathbb{I}^{1,1}_{k}$ can be equal to zero; however, capture-recapture studies where no one gets captured twice are trivial and we therefore exclude this possibility from our considerations. Now, plugging (5) into Jensen's equality for a lower bound,
\begin{center}
$\frac {z^2 S_1 S_2}{N\alpha_1 \alpha_2 m_2} \leq\mathbb{E}[\frac {S_1 S_2} {\sum_{k}\mathbb{I}^{1,1}_{k}}] $
\end{center}
and into Kantorovich's inequality for an upper bound
\begin{center}
$\mathbb{E}[\frac {S_1 S_2} {\sum_{k}\mathbb{I}^{1,1}_{k}}] \leq \frac{(S+1)^2}{4S} \frac {z^2 S_1 S_2}{N\alpha_1 \alpha_2 m_2} $
\end{center}
with $S$ denoting $min(S_1,S_2)$. Combined and simplified, the two give
\begin{equation}
\frac {z^2}{m_2} N\leq \mathbb{E}[\frac {S_1 S_2} {\sum_{k}\mathbb{I}^{1,1}_{k}}] \leq \frac{(S+1)^2}{4S} \frac {z^2}{m_2} N
\end{equation}
Thus we conclude that for a homogeneous catchability (i.e., a degree distribution  with $m_2\approx z^2$) a researcher using a naive LP estimator, thus worried about an underestimate, can be reassured by the lower bound which is close to $N$.
Quite the opposite  is true for heterogeneous networks (having relatively large $m_2$) in that the correction terms to $N$ might take very small values, yielding  a gross underestimate (see figure 1).

\subsection{Convergence and robustness of $\hat{N}_{adj}$}
Turning to $\hat{N}_{adj}$, our first main result is that asymptotically $\hat{N}_{adj}$ converges in probability to the true population size. The strategy for proving this is to notice that both in the numerator of the right hand side of (3), as well as in the denominator, we have a sum of Bernoulli-like random variables. Since a sum of such variables is concentrated close to its expectation (as we show in the appendix using the method of bounded differences (lemma 1)) we can find the expectation of the expression and conclude that their ratio is also concentrated close to what we need.

In the process proving of theorem 1 we also find our second main result, that as long as we use the appropriate capture probabilities for one of the sampling stages, it does not matter what the capture probabilities are for the other stage.

Let $\pi_i := \frac{S_2}{N}\tilde{\pi}_i$ be the probability to sample individual $i$ at the recapture stage with a sample size of $S_2$, with the $\tilde{\pi}_i$'s being constants associated to an observable covariate. Instead of (3), we can (and should, from a researcher running the survey's point of view) use the $\tilde{\pi}_i$'s and be concerned with
\begin{equation}
\hat{N}_{adj} = (\sum_{k}{\mathbb{I}^{1,0}_{k}}+\sum_{k}{\mathbb{I}^{1,1}_{k}})\frac{(\sum_{k} \frac{\mathbb{I}^{1,1}_{k}}{\tilde{\pi}_k}+\sum_{k}\frac{\mathbb{I}^{0,1}_{k}}{\tilde{\pi}_k})}{\sum_{k}\frac{\mathbb{I}^{1,1}_{k}}{\tilde{\pi}_k}}
\end{equation}

And now we have
\\
\\
\noindent \emph{Theorem 1. As} $N\rightarrow \infty$, $\frac{\hat{N}_{adj}}{N} \overset{P}{\longrightarrow}1$
\\
\\
\noindent \emph{Proof. See appendix.}
\\
\\
\noindent Remark 1.: In general, we do not have convergence in expectation, i.e., $\mathbb{E}[\frac{\hat{N}_{adj}}{N}] \rightarrow 1$. However, if we restrict ourselves to cases where at least one individual is captured twice (and thus $\hat{N}_{adj}$ is bounded from above by $N^2$) we have  $\mathbb{E}[\frac{\hat{N}_{adj}}{N}] \rightarrow 1$ as well.
\\
\\
\noindent Remark 2.: As a by-product, note that in the proof of the theorem 1 we made no assumptions about the capture probabilities in the first stage, apart from the fact that they sum up to $\alpha_1 N$; thus we also have
\\
\\
\noindent \emph{Theorem 2.} $\hat{N}_{adj}$ \emph{is indifferent to changed capture probabilities in the first stage}.

\subsection{Simulations}

To demonstrate the sensitivity of the naive estimator to heterogeneity in the capture probabilities and compare it to $\hat{N}_{adj}$, we simulated a capture-recapture RDS-like design where the probability of sampling is correlated with
the capture and recapture stages, because the probability of sampling individual $i$ is proportional to his degree $d_i$. We calculated the naive and adjusted estimators for a population with a power-law degree distribution, over a range of values of
$\lambda$, which is a measure of the heterogeneity in degree between individuals. For each $\lambda$,
we sampled the population twice; in the capture phase, individuals were sampled in relation to their degree; and in the recapture
phase, individuals were sampled by an RDS-like process, and $\hat{N}_{naive}$ and $\hat{N}_{adj}$ were
calculated (figure 1). For large values of $\lambda$, corresponding to homogenous degrees, $\hat{N}_{naive}$ and $\hat{N}_{adj}$ were similar and close to the true population size. However, as $\lambda$ decreased, the naive estimate decreased and
worsened, while $\hat{N}_{adj}$ was relatively constant across the range of $\lambda$. If individuals have equal capture probabilities (but recapture is by RDS), then $\hat{N}_{naive}$ and $\hat{N}_{adj}$ are similar, irrespective of $\lambda$ (not shown).

\begin{figure}\centerline{\includegraphics[width=10cm]{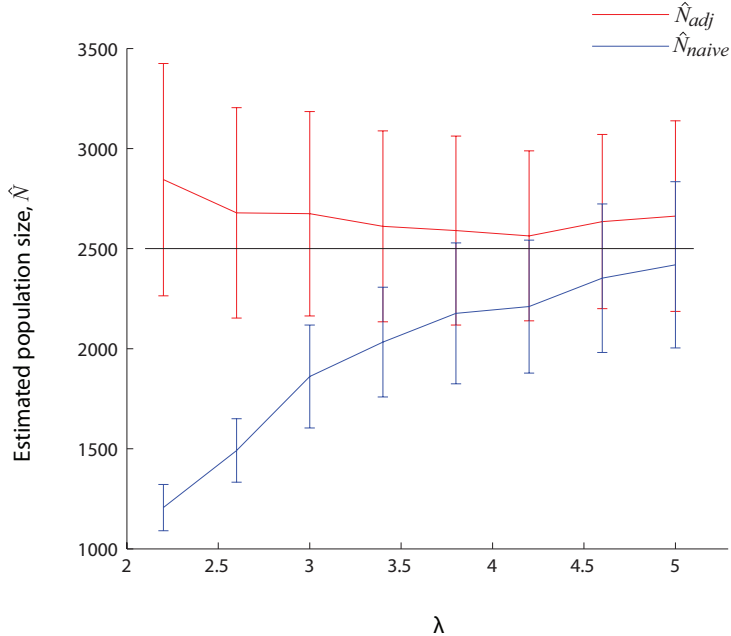}}
\caption{Naive and adjusted estimates for a population with a power-law degree distribution, over a range of values of degree heterogeneity $(\lambda)$. Mean and one standard deviation of naive (blue) and adjusted (red) estimates of population size across a range of networks with different levels of degree heterogeneity captured by the parameter $\lambda$, with high values corresponding to networks with a
relatively uniform degree, assuming sampling proportional to degree during the capture stage. A total population size of $N=2500$ and equal numbers of individuals in the capture and recapture stages, $S_1=S_2=100$, were used. Random networks were
generated according to a power-law distribution, $p_n\sim n^{-\lambda}$, with a minimum degree of 3. The recapture phase was simulated by an RDS-like process, in which a single 'seed' individual was sampled, and one of its network neighbours sampled; this is followed by random sampling of the
combined neighbours of these two nodes and so on, until the desired sample size was obtained. To avoid recruitment rates being strongly affected by degree, we limited the number of recruits per individual to 3. For each value of $\lambda$, 20 networks were constructed and on each one, the process
was run 50 times (i.e., 1000 repetitions per $\lambda$).  }
\end{figure}

\section{Evaluation on empirical data}

In this section we examine the theory on different data-sets, focusing on a survey from rural Uganda (sec. 3.1) which is the most comprehensive, but also on two others \cite{Goel2010,Wejnert2009}.

This section can also be considered an examination of RDS sampling and inference. Accordingly, we mostly use different empirically obtained RDS trees (and combinations of them) as a recapture stage. Starting with the Uganda data-set, we first use simple random (uniform) sampling or degree-biased sampling from the general population as a capture stage (section 3.1). Then use records of association to an \textquotedblleft age" category as the \textquotedblleft capture" group.

Section 3.2 wraps up with an evaluation of the method on the P90 network, recently used for an examination of the utility of RDS \cite{Goel2010}, and on a WebRDS survey from a large university \cite{Wejnert2009}. Using
official institutional records of the number of students from different racial groups we were able to use them as a \textquotedblleft capture" stage (similar to using the \textquotedblleft age" data in sec. 3.1), thus implementing the method in a simple manner.

\subsection{Evaluation on the Uganda data-set}

The data used to define our first \textquotedblleft target population" were available from an ongoing general population cohort of 25 villages in rural Masaka, Uganda.  Annually, after obtaining consent, a total-population household census and an individual questionnaire are administered and blood taken for HIV-1 testing from this population.
The target population consisted of $N=2402$ men who were recorded as a male head of a household within the general population cohort study villages between February 2009 and January 2010.

As part of an evaluation of RDS methodology
an RDS survey was carried out on this population, employing current RDS methods of sampling and statistical inference (details are available in full in \cite{McCreesh}). Briefly, 10 seeds recruited 927 male household heads over 54 days using RDS.  The total number of recruits originating from each seed ranged from 8 to 241 (0.9\% to 26.3\%). The number of waves ranged from 3 to 16.  The mean degree of RDS recruits (including seeds) was 12.1. Data were also collected on a simple random sample of 300 eligible male household heads who were not recruited during the RDS. 54\% (162/300) completed the interview.  The distribution of the reported degree of RDS recruits  was approximately Gaussian and showed likely over-reporting of multiples of 5 (fig 2), similarly to the SRS sample of non-recruits.

\begin{figure}\centerline{\includegraphics[width=15cm]{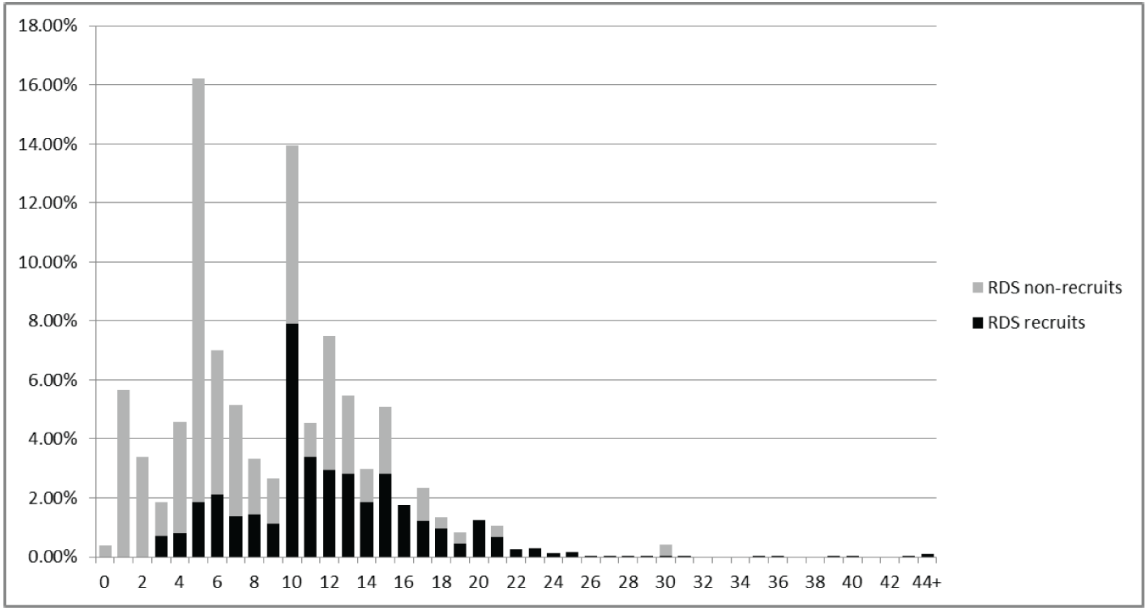}}
\caption{The distribution of the reported degree of RDS and SRS recruits. }
\end{figure}

Thanks to the extensive nature of the Uganda data-set we were able to assign degrees to all individuals in the population using the degrees in the simple random sample (SRS). This was done by concatenating the degree sequence in the SRS (162 individuals) until a degree sequence of length 1475 (=2402-927) was obtained.

For our first experiment we sampled randomly from the general population to obtain a \textquotedblleft capture" group and used the RDS data to obtain a \textquotedblleft recapture" group. For the left column of figure 3 (panels a,c,e and g) we used a simple random (uniform) sample of 250 individuals from the general population as the capture stage; whereas for the right column of figure 3 (panels b,d,f and h) we used a degree-biased sample of 250 individuals from the general population. Having 10 different RDS trees there is a total of $2^{10}-1=1023$ different combinations of trees possible as the recapture group. In addition, it is also possible to \textquotedblleft bootstrap" and sample uniformly at random from the 927 RDS participants, a subsample of, say, 200 individuals. In order to preserve clarity, figure 3 shows only the results of using combinations of 1,2,8 or 9 RDS trees as the recaptured group. For each possible combination of a single tree (top two panels a and b), a pair of trees (panels c and d), eight trees (panels e and f) or nine (panels g and h) we evaluated $\hat{N}_{naive}$ (blue circles) and $\hat{N}_{adj}$ (red squares). 50 repetitions were done (of sampling from the general population) and the mean and an error bar of one standard deviation are plotted vs the size of the combined group (i.e., number of individuals). Note that, as expected, when sampling was done uniformly at random (a,c,e,g) both estimators yielded a good estimate; however,  when sampling was done in a degree-biased manner (b,d,f,h) the naive estimator substantially underestimated the size of the population whereas the adjusted estimator still provided a good estimate. Using other combinations of trees gave similar results (data not shown).

In addition, we also tested the estimator with a \textquotedblleft bootstrap" recapture group selected (uniformly) at random from the RDS population, with the result shown in figure 4. It should be noted that we did not sample this recapture group (RDS subsample) according to its degree, and therefore it is very encouraging to see that weighting by the \emph{reported} degrees improved the estimates almost to perfection.

\begin{figure}\centerline{\includegraphics[width=15cm]{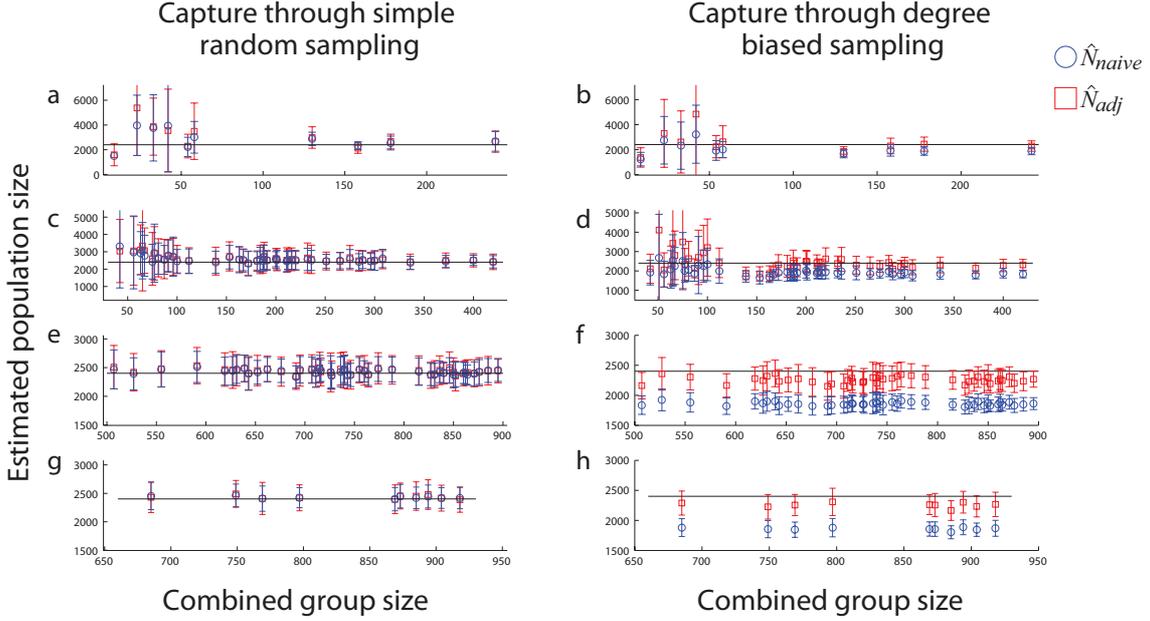}}
\caption{Estimated population size for different combinations of RDS trees used as the \textquotedblleft recapture" sample (vs the size of the combined group).
For the left column of figure 3 (panels a,c,e and g) we used a simple random (uniform) sample of 250 individuals from the general population as the capture stage; whereas for the right column of figure 3 (panels b,d,f and h) we used a degree-biased sample of 250 individuals from the general population. In order to preserve clarity, we only shows the results of combinations of 1,2,8 or 9 RDS trees as the recaptured group. For each possible combination of a single tree (top two panels, a and b), a pair of trees (panels c and d), eight (panels e and f) or nine trees (panels g and h) we evaluated $\hat{N}_{naive}$ (blue circles) and $\hat{N}_{adj}$ (red squares). 50 repetitions were done (of sampling from the general population) and the mean and an error bar of one standard deviation are plotted vs the size of the combined group (i.e., number of individuals)
 Black line shows true population size (2402). }
\end{figure}

\begin{figure}\centerline{\includegraphics[width=10cm]{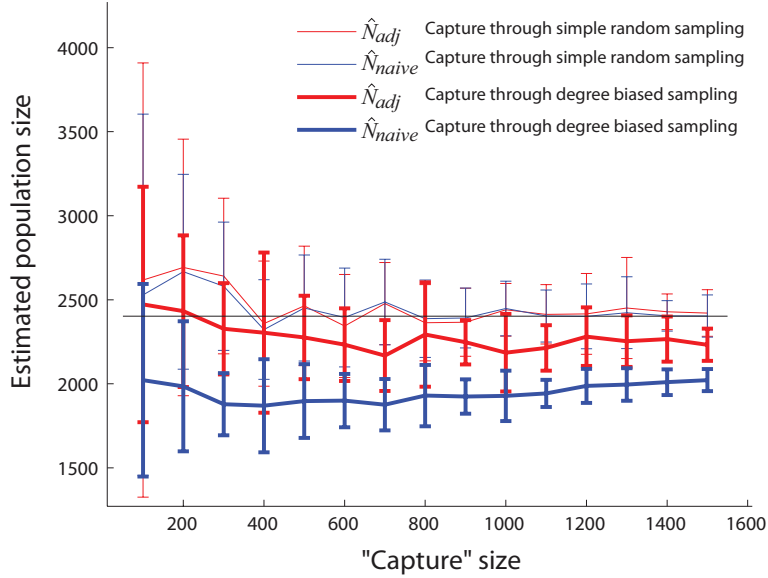}}
\caption{Estimated population size vs the size of the "capture" sample. After sampling $X$ individuals (where $X=100,200,...,1500$) from the general population uniformly random (thin lines) or with probability proportional to their degree  (thick lines) we sampled 200 individuals from the 927 RDS participants (\emph{uniformly} random) as the \textquotedblleft recapture" sample. For each of the 15 cases we performed 50 repetitions and plotted the mean of the estimator with error bars representing the standard deviation.  Notice $\hat{N}_{adj}$ (red) is robust with respect to the catchabilities in the capture stage (theorem 2); i.e. it works well for both sampling methods.  }
\end{figure}

In many surveys, and in RDS in particular, there are rarely two different  capture stages. In fact, it could be argued that the \textquotedblleft capture" stage in our simulations is not easily, or cheaply, reproducible in practice. We therefore took a more realistic and challenging approach:
using records of association to an \textquotedblleft age" category as the \textquotedblleft capture" group we reproduced the above analysis. It should be noted, as discussed further in section 4, that this approach is not only highly generalizable
but also does not involve violation of the privacy of individuals not participating in the \textquotedblleft recapture" stage; the only information required from the \textquotedblleft records" is the number of recorded individuals.

Each age group at a time (0-19, 20-29, 30-39, 40-49 and 50+, with sizes 47, 484, 660, 496 and 714 respectively) was used as a \textquotedblleft capture" stage. 300 individuals were then  sampled uniformly at random from the RDS sample as a \textquotedblleft recapture" stage and estimates of $N$ were found; 100 repetitions were done for each age group and the mean and one standard deviation, $\sigma$, are plotted vs the size of each group (fig.5). The large overestimate based on the 0-19 age group data (with its large variability; $\sigma_{naive}= 3231$ and $\sigma_{adj}= 6402$) is probably not only due to the small size of the group (see, e.g., the large variance in fig. 4 for small \textquotedblleft capture" group size), but rather, as an additional qualitative survey revealed \cite{McCreesh}, because young unmarried \textquotedblleft head of households" were not considered eligible for recruitment by their peers, and thus were undersampled. Estimates from the rest of the age groups are reasonable, surrounding $N=2402$, and when averaged over all such age groups yielded an overestimate of $\sim 10\%$ relative to $N=2402$.

The relatively similar behavior of the both estimators can be attributed to the fact that the degree distribution is rather homogenous (cf. fig. 6, described in sec. 3.2, obtained from a network with a more heterogenous degree distribution), and, of course, by the fact that different age groups might be less strongly correlated with degree than what one might presume.


\begin{figure}\centerline{\includegraphics[width=7cm]{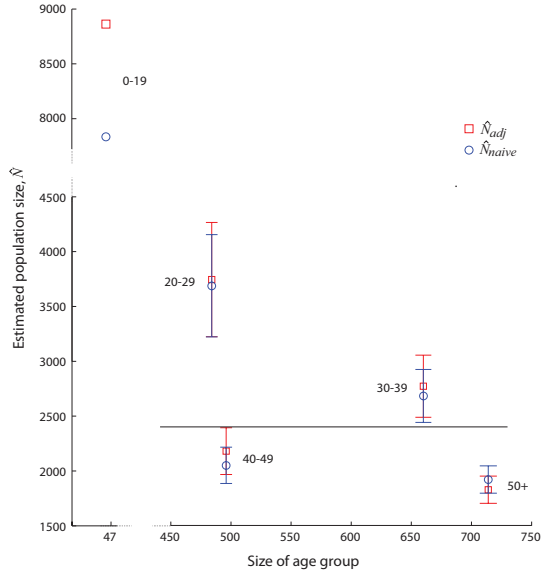}}
\caption{Estimating the size of the population using the \textquotedblleft age" category as a \textquotedblleft capture" stage. Each of the age groups 0-19, 20-29, 30-39, 40-49, 50+ (with true population sizes of 47, 484, 660, 496, 714 respectively) at a time was used as a \textquotedblleft capture" stage. 300 individuals were then  sampled uniformly at random from the RDS sample as a \textquotedblleft recapture" stage and estimates of $N$ were found; 100 repetitions were done and the mean and one standard deviation, $\sigma$, are plotted vs the size of each group. The large overestimate from the 0-19 age group (with its large variability; $\sigma_{naive}= 3231$ and $\sigma_{adj}= 6402$) is probably not only due to the small size of the group (see, e.g., the large variance in fig. 4 for small \textquotedblleft capture" group size), but rather because young unmarried \textquotedblleft head of households" were under-recruited (see text).  }
\end{figure}


\begin{figure}\centerline{\includegraphics[width=7cm]{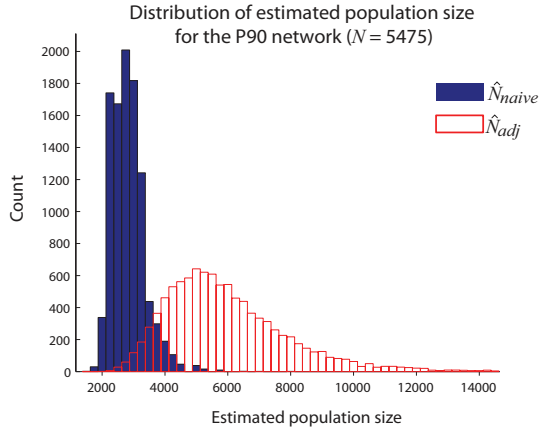}}
\caption{The distribution of $\hat{N}_{adj}$ and $\hat{N}_{naive}$ for Project 90. After simulating an RDS \textquotedblleft capture" stage, as in sec 2.2, with 300 individuals, we sampled 300 individuals again as the \textquotedblleft recapture" stage and evaluated $\hat{N}_{adj}$ and $\hat{N}_{naive}$. This was done $10^4$ times and the distributions of $\hat{N}_{adj}$ and $\hat{N}_{naive}$ were plotted.}
\end{figure}

\subsection{Additional data sets - Project 90 and WebRDS}
In addition to testing the method on the Uganda data, we also evaluated it on two additional data-sets. Although not as extensive as the Uganda data, they complement each other in a sense - the first, Project 90, consists of a detailed network (without an RDS sample), whereas the second, the WebRDS data, is made up of a web-based respondent driven sample of students of a large university (without details of the complete network).

The first source of data, Project 90, was a large study that began in 1987 that was designed to examine the influence of network structure on the propagation of infectious disease by constructing a network census ($N=5475$) of high-risk heterosexuals in Colorado Springs \cite{Klovdahl}. Recently, the Project 90 network was used to assess the utility of RDS \cite{Goel2010} ; we therefore used it as our networked population, and attempted to infer its size by RDS and the adjusted estimator. After simulating an RDS \textquotedblleft capture" stage, as in sec 2.2, with 300 individuals, we sampled 300 individuals again as the \textquotedblleft recapture" stage and evaluated $\hat{N}_{adj}$ and $\hat{N}_{naive}$. This was done $10^4$ times and the distributions of $\hat{N}_{adj}$ and $\hat{N}_{naive}$ were obtained (fig. 6). Note that, in agreement with the findings of ref. \cite{Goel2010}, although an RDS adjusted estimator has a small bias ($N=5475$, average $\hat{N}_{adj}=5970$), it does have a larger variance
than the naive estimator.
However, fig. 6 clearly  shows that the adjusted estimator is much better concentrated near the true population size.

The second source of data, the university WebRDS  \cite{Wejnert2009}, is much smaller, with data available only for the 378 participants in the sample. However, this sample, genuinely driven the respondents themselves, together with available official institutional records \cite{Cornell08}, allowed us to conduct a proper examination of both estimators.

First, we used the race group \textquotedblleft White" as a capture group, with size $=6716$ according to institutional records, and checked how many white students (and with which degrees) appeared in the WebRDS sample, thus finding a way to obtain $\hat{N}_{adj}=11055$ and $\hat{N}_{naive}=9437$. Then, we performed the same procedure with the rest of the racial tags (table 1). While the large overestimation using the \textquotedblleft Hispanic" group looks discouraging, a possible explanation is that only 8 Hispanics were \textquotedblleft recaptured", probably because the initial convenience sample had only white (7) black (1) and Asian (1) seeds.
When \textquotedblleft capturing" other race groups, most of estimates were much closer to the true population size ($N=13510$ according to institutional records), and with a slight advantage to $\hat{N}_{adj}$. This advantage is further strengthened when taking into consideration various problems in the data due to incomplete records, which led ref. \cite{Wejnert2009} to suggest that $N\approx 11750$ should be considered as the actual population size of known racial affiliation.

\begin{table}
  \centering
  \begin{tabular}{|c|c|c|c|c|c|}
    \hline
    Race & White & Asian & Hispanic & Black & non US \\
    Enrolled & 6716 & 2191 & 748  &  712 & 1073 \\
    In sample & 269 & 58 & 8 & 14 & 14 \\
    $N_{naive}$ &  9437 & 14279 & 35343 & 19224 &  28971 \\
    $N_{adj}$ &  11055 & 10728 & 39482  & 13806 &  20552 \\
    \hline
  \end{tabular}
  \caption{University statistics: estimating the number of students using racial groups as a \textquotedblleft capture" stage. $N=13510$ in institutional records, effective $N\approx 11750$ \cite{Wejnert2009}, see text.}
\end{table}

\section{Discussion}

Here we considered capture-recapture experiments with heterogenous catchability. In contrast to common practice, we consider (unknown) capture probabilities that depend on the size of the population (and thus many might take small values), while on the other hand we allow the assumption that they are related to an observed covariate in some manner.

An important example for such setting, and the application we had in mind, is respondent driven sampling, where the probability to sample an individual is assumed to be proportional to his degree.
In addition to the concern whether the degree of an individual has a direct bearing on its probability of being sampled, there are at least two other concerns  regarding this assumption implicit in application of RDS:

\noindent first, sampled individuals might incorrectly report their own degree, possibly causing some bias. Second, as noted in \cite{Gile11}, if a large fraction of the population is sampled there will be a deviation from proportionality between degree and catchability.

Indeed, errors in assessing one's own degree can be addressed by using, e.g., the scale-up method \cite{ScaleUp}, and deviations from proportionality could be approached by a method in the spirit of \cite{Gile11}. However, deviations from proportionality are generally only problematic after a large fraction has been sampled; indeed the simulations in \cite{Gile11} have a nice capture-recapture interpretation: after marking $20\%$ of the population (with some preference according to degree) a subsequent RDS-like sample was obtained; even when $50\%$ of the population was recaptured, a simple RDS weighting yielded a good estimate of the fraction of marked individuals.

Furthermore, although such corrections might be considered and applied, there is evidence suggesting this is not crucial; our analysis of the Uganda data, for example, gave encouragingly small bias when using the \emph{reported} degrees, even for a sample of size 927/2402.

Perhaps the most interesting experiment reported here involves using the \textquotedblleft age" and \textquotedblleft race" categories when estimating the size of the population (for, respectively, the Uganda and the WebRDS data-sets). While the idea of using various patient records or lists in \textquotedblleft capture-recapture" studies is not new, it is usually employed for ascertaining the number of people with a certain medical condition (related directly to being on such a list) and not the size of the general population at risk (see \cite{Chao} and references therein). On the other hand, the WHO/UNAIDS `Guideline on Estimating the Size of Populations Most at Risk to HIV' \cite{WHO2010}, for example, embraces the use of capture-recapture methods but strongly advocates against violating the assumptions of the simple model (i.e., homogenous catchability, and excluding \textquotedblleft non-random" incidence lists). Nevertheless, as suggested from theorem 2 and demonstrated in the experiment in sec. 3, even if the first capture stage is non-random we should still be able to obtain a good estimate. It is also worth noting there is no violation of individual's privacy, since the only information required from the records is the number of registered individuals. Thus, the method suggested here could easily be adapted and used for other important applications in many diverse scenarios utilizing a \textquotedblleft capture" stage, as in sec. 3.2 or as the \textquotedblleft key-chain" method \cite{PB}, where there is no \textit{a-priori} guarantee individuals will be sampled randomly in a uniform manner.

\section{Appendix}
Theorems 1 and 2  are derived in detail below.

Recall the following inequality by McDiarmid \cite{McDiarmid}:
\\
\\
\noindent \emph{Lemma 1.} [McDiarmid]\emph{: Let} $X_1,X_2,...,X_N$ \emph{be real valued random variable and let} $\overrightarrow{X}$ \emph{denote the random vector }$[X_1,X_2,...,X_N]$. \emph{Consider a function} $f:R^N\rightarrow R$ \emph{and suppose there exist a constant,} $c$\emph{, such that}
\begin{center}
$|\mathbb{E}[f(\overrightarrow{X})\mid [X_1,X_2,...,X_{k-1}]=[x_1,x_2,...,x_{k-1}], X_k = x_k]$\\$-\mathbb{E}[f(\overrightarrow{X})\mid [X_1,X_2,...,X_{k-1}]=[x_1,x_2,...,x_{k-1}], X_k = x'_k]|\leq c$
\end{center}
\emph{for each} $k=1,2...,N$ \emph{and} $x_i$ (i=1,2,...,k)\emph{. Then for any }$t$
\begin{center}
$\mathbb{P}(|f(\overrightarrow{X})- \mathbb{E}[f(\overrightarrow{X})]| \geq t)\leq 2exp[\frac{-2t^2}{Nc^2}]$
\end{center}

Denote by $\mathbb{I}^{x,1}_{k}$  the indicator function whether individual $k$ was recaptured at the second stage, regardless of the first stage. Considering the sequence of $N$ random variables,  $\{X_k\}_{k=1}^N$, with $X_k:=\frac{\mathbb{I}^{x,1}_{k}}{\tilde{\pi}_k}$, and the function $f(\overrightarrow{X})= \sum_{k=1}^N X_k$ we have
\begin{equation}
\mathbb{E}[f(\overrightarrow{X})]
=\mathbb{E}[\sum_{k=1}^N X_k]
=\mathbb{E}[\sum_{k=1}^N \frac{\mathbb{I}^{x,1}_{k}}{\tilde{\pi}_k}]
= \sum_{k=1}^N \pi_k\frac{1}{\tilde{\pi}_k}
= \sum_{k=1}^N \frac{S_2}{N}\tilde{\pi}_k\frac{1}{\tilde{\pi}_k}=S_2=\alpha_2N
\end{equation}

Without loss of generality, we can assume that $min_i(\tilde{\pi}_i)=1$ and has a mutual bound $c=1$, as required by lemma 1, and obtain
\\
\\
\noindent \emph{Lemma 2.} $\sum_{k=1}^N \frac{\mathbb{I}^{x,1}_{k}}{\tilde{\pi}_k}$ \emph{is concentrated near} $\alpha_2 N$ \emph{with}
\begin{center}
$\mathbb{P}(|\sum_{k=1}^N \frac{\mathbb{I}^{x,1}_{k}}{\tilde{\pi}_k}- \alpha_2 N| \geq \sqrt{N\cdot logN})\leq \frac{2}{N^2}$
\end{center}
\noindent \emph{Proof. Apply lemma 1, with} $f$ \emph{and} $c$ \emph{as above.}
\\
\\
Similarly, denoting by $\beta_i$'s the capture probabilities in the first stage, and using only the fact that they sum up to $\alpha_1 N$ we have
\begin{equation}
\mathbb{E}[\sum_{k=1}^N \frac{\mathbb{I}^{1,1}_{k}}{\tilde{\pi}_k}]
= \sum_{k=1}^N \beta_k \pi_k\frac{1}{\tilde{\pi}_k}
= \sum_{k=1}^N \beta_k \frac{S_2}{N}\tilde{\pi}_k\frac{1}{\tilde{\pi}_k}=\alpha_1\alpha_2N
\end{equation}
and obtain
\\
\\
\noindent \emph{Lemma 3.} $\sum_{k=1}^N \frac{\mathbb{I}^{1,1}_{k}}{\tilde{\pi}_k}$ \emph{is concentrated near} $\alpha_1 \alpha_2 N$ \emph{with}
\begin{center}
$\mathbb{P}(|\sum_{k=1}^N \frac{\mathbb{I}^{1,1}_{k}}{\tilde{\pi}_k}- \alpha_1\alpha_2 N| \geq \sqrt{N\cdot logN})\leq \frac{2}{N^2}$
\end{center}
\noindent \emph{Proof. The same as for lemma 2 above.}
\\
\\
Our first main result is:
\\
\\
\noindent \emph{Theorem 1. As} $N\rightarrow \infty$, $\frac{\hat{N}_{adj}}{N} \overset{P}{\longrightarrow}1$
\\
\emph{(a) with}
\begin{center}
$\mathbb{P}( \frac{\alpha_1 \alpha_2 N - \alpha_1\sqrt{N\cdot logN}}{\alpha_1 \alpha_2 N + \sqrt{N\cdot logN}}
\leq\frac{\hat{N}_{adj}}{N}
\leq  \frac{\alpha_1 \alpha_2 N + \alpha_1 \sqrt{N\cdot logN}}{\alpha_1 \alpha_2 N - \sqrt{N\cdot logN}})
\geq 1 - \frac{4}{N^2}$
\end{center}
\emph{(b) and in expectation. I.e.,} $\mathbb{E}[\frac{\hat{N}_{adj}}{N}] \rightarrow 1$.
\\
\\
\noindent \emph{Proof. (a) is obtained via a simple combination of lemmata 3 and 4. As for (b), recall we have at least one individual captured twice and thus $\hat{N}_{adj}$ is bounded from above by $N^2$. Now (b) follows from (a) easily.}
\\
\\
Noticing we have made no assumptions about the $\beta_i$'s, the capture probabilities in the first stage, apart from the fact that they sum up to $\alpha_1 N$, we also have
\\
\\
\noindent \emph{Theorem 2.} $\hat{N}_{adj}$ \emph{is robust with respect to different choices of the the capture probabilities in the first stage,} $\{\beta_i\}_{i=1}^N$.


\newpage
\bibliographystyle{plain}
\bibliography{Ucrc}

\end{document}